\documentclass[12pt, prd, showpacs]{revtex4}
\usepackage{amssymb}
\usepackage{amsmath}

\input{tcilatex}

\begin{document}

\title{High energy particle collisions and geometry of horizon}
\author{O. B. Zaslavskii}
\affiliation{Department of Physics and Technology, Kharkov V.N. Karazin National
University, 4 Svoboda Square, Kharkov 61022, Ukraine}
\affiliation{Institute of Mathematics and Mechanics, Kazan Federal University, 18
Kremlyovskaya St., Kazan 420008, Russia}
\email{zaslav@ukr.net }

\begin{abstract}
We consider collision of two geodesic particles near the lightlike surface
(black hole horizon or naked singularity) of such an axially symmetric
rotating or static metric that the coefficient $g_{\phi \phi }\rightarrow 0$
on this surface. It is shown that the energy in the centre of mass frame $%
E_{c.m.}$ is indefinitely large even without fine-tuning of particles'
parameters. Kinematically, this is collision between two rapid particles
that approach the horizon almost with the speed of light but at different
angles (or they align along the normal to the horizon too slowly). The
latter is the reason why the relative velocity tends to that of light, hence
to high $E_{c.m.}$. Our approach is model-independent. It relies on general
properties of geometry and is insensitive to the details of material source
that supports the geometries of the type under consideration. For several
particular models (the stringy black hole, the Brans-Dicke analogue of the
Schwarzschild metric and the Janis-Newman-Winicour one) we recover the
results found in literature previously.
\end{abstract}

\maketitle

\section{Introduction}

Several years ago, an interesting observation was made by Ba\~{n}ados, Silk
and West (the BSW effect, after the names of its authors). It turned out
that if \ two particles collide near the black hole horizon, their energy in
the centre of mass frame $E_{c.m.}$ can grow unbounded \cite{ban}. At first,
this was obtained for the Kerr black hole but later on, it was shown that
this is a generic feature of rotating black holes \cite{prd}. Typically, $%
E_{c.m.}$ remains modest. It grows unbounded under special conditions only.
Namely, one of particles should have special relation between the energy and
the angular momentum (so-called critical particle), whereas another one
should be not fine-tuned (so-called usual) - see aforementioned papers for
details.

Meanwhile, several papers appeared in which it was found that unbounded $%
E_{c.m.}$ can be obtained without fine-tuning at all! As there is a sharp
contrast (or even seeming contradiction) between these results and the
standard picture described in the first paragraph, special explanation is
needed here. In \cite{jnw}, unbounded $E_{c.m.}$ were obtained for the naked
singularities described by the Janis-Newman-Winicour metric. In \cite{fern},
this was obtained for stringy black holes and in \cite{s2} \ for the
Brans-Dicke analogue of the Kerr black hole (so-called BDK metric \cite{kim}%
). Thus very different metrics and completely different types of spacetime
(naked singularities and black holes) give the same results and this appeals
to explanation.

In the present paper, we consider "dirty" black holes (see, e.g. \cite{dirty}%
) and develop a general approach that handles such cases independently of
the details of a metric and material source that supports it. This approach
agrees with previous particular results. The type of spacetime under
consideration extends the set of geometries for which the high energy
collisions are possible.

Throughout the paper, we put fundamental constants $G=c=1$.

\section{Metric, equations of motion}

We consider the general metric of the form

\begin{equation}
ds^{2}=-N^{2}dt^{2}+g_{\phi }(d\phi -\Omega dt)^{2}+\frac{dr^{2}}{A}%
+g_{\theta }d\theta ^{2}  \label{met}
\end{equation}%
in which all metric coefficients do not depend on $t~\,$and $\phi $.
Therefore, the energy $E=-mu_{0}$ and the angular momentum $L=mu_{\phi }$
are conserved, where $m$ is the particle's mass, $u^{\mu }=\frac{dx^{\mu }}{%
d\tau }$ is the four-velocity, $\tau $ is the proper time. In what follows,
we consider motion within the equatorial plane $\theta =\frac{\pi }{2}$
only. Then, we redefine the radial coordinate in such a way that $A=N^{2}$.
The equations of motion for geodesics read%
\begin{equation}
m\dot{t}=\frac{X}{N^{2}}\text{,}  \label{0}
\end{equation}%
\begin{equation}
X=E-\Omega L\text{,}  \label{X}
\end{equation}%
\begin{equation}
m\dot{\phi}=\frac{L}{g_{\phi }}+\frac{\Omega X}{N^{2}}\text{,}  \label{phi}
\end{equation}%
\begin{equation}
m\dot{r}=\sigma Z\text{,}  \label{r}
\end{equation}%
where $\sigma =\pm 1$ depending on the direction of radial motion,%
\begin{equation}
Z=\sqrt{X^{2}-N^{2}(m^{2}+\frac{L^{2}}{g_{\phi }})}.  \label{z}
\end{equation}

\section{General formulas for collision}

Let two particles 1 and 2 collide. The energy in their centre of mass frame
is equal to%
\begin{equation}
E_{c.m.}^{2}=-(m_{1}u_{1\mu }+m_{2}u_{2\mu })(m_{1}u_{1}^{\mu
}+m_{2}u_{2}^{\mu })=m_{1}^{2}+m_{2}^{2}+2m_{1}m_{2}\gamma \text{,}
\label{cm}
\end{equation}%
where 
\begin{equation}
\gamma =-u_{1\mu }u^{2\mu }  \label{gau}
\end{equation}%
is the Lorentz gamma factor of relative motion.

We assume that in (\ref{r}) $\sigma _{1}=\sigma _{2}=-1$ that is typical of
particle motion near black holes. Then, it follows from (\ref{0}) - (\ref{z}%
) that 
\begin{equation}
m_{1}m_{2}\gamma =\frac{X_{1}X_{2}-Z_{1}Z_{2}}{N^{2}}-\frac{L_{1}L_{2}}{%
g_{\phi }}.  \label{ga}
\end{equation}

\section{Near-horizon collisions of usual particles}

If the black hole horizon exists, it is located at $N=0$. We include into
consideration also singularities for which $N=0$. Although, strictly
speaking, such a surface cannot be called a horizon, we will use for brevity
the term "horizon" both in the regular and singular cases. According to \cite%
{ban} and its generalization in \cite{prd}, if collision occurs at the point
with small $N,$ the factor $\gamma $ can be indefinitely large, provided one
of particles is fine-tuned ("critical") whereas another one ("usual") is
not. In doing so, it was tacitly assumed that $g_{\phi }$ remains nonzero.

But let us consider another case, when simultaneously $N\rightarrow 0~$and 
\begin{equation}
g_{\phi }\rightarrow 0.  \label{gf}
\end{equation}%
We will be interested in collision of usual (not fine-tuned) particles. We
denote $\frac{N^{2}}{g_{\phi }}=b^{2}$ . Then, for collision at the point
with small $N$, we have%
\begin{equation}
m_{1}m_{2}\gamma \approx \frac{D}{N^{2}}\text{ },  \label{gd}
\end{equation}%
\begin{equation}
D=X_{1}X_{2}-\sqrt{X_{1}^{2}-L_{1}^{2}b^{2}}\sqrt{X_{2}^{2}-L_{2}^{2}b^{2}}%
-b^{2}L_{1}L_{2}\text{.}  \label{D}
\end{equation}%
It can be also rewritten in the form%
\begin{equation}
m_{1}m_{2}\gamma \approx \frac{C}{g_{\phi }}\text{, }C=\frac{D}{b^{2}}\text{,%
}  \label{gc}
\end{equation}%
useful for small $b$ (see below).

Introducing $\alpha _{i}$ ($i=1,2)$ according to%
\begin{equation}
bL_{i}\equiv X_{i}\sin \alpha _{i}\text{, }-\frac{\pi }{2}\leq \alpha
_{i}\leq \frac{\pi }{2}\text{,}  \label{xa}
\end{equation}%
we can rewrite (\ref{gd}) as%
\begin{equation}
m_{1}m_{2}\gamma \approx \frac{X_{1}X_{2}d}{N^{2}}\text{ },  \label{xd}
\end{equation}%
\begin{equation}
d=1-\cos (\alpha _{1}-\alpha _{2})\text{.}  \label{d}
\end{equation}

Now, we consider two different limiting cases depending on the near-horizon
behavior of $b$.

\subsection{$b\rightarrow 0$}

Then, $D=O(b^{2})$ but $C$ is, in general, nonzero, 
\begin{equation}
C=\frac{(L_{1}X_{2}-L_{2}X_{1})^{2}}{2X_{1}X_{2}}.  \label{c}
\end{equation}

If $g_{\phi }$ is finite on the horizon, the quantity $\gamma $ is finite as
well according to (\ref{gc}). This is in accord with the fact that collision
of two usual particles cannot produce unbounded $E_{c.m.}$ \cite{prd}.
However, for extremely small $g_{\phi }$, this \ rule is not valid anymore
and, according to (\ref{gc}), we obtain indefinitely large $\gamma $.

\subsection{$b\rightarrow \infty $}

For nonzero $L_{1,2}$ this is impossible since it is seen from (\ref{z})
that this violates the condition $Z^{2}>0$. However, we can adjust $L_{1}$
and $L_{2}$ to the big $b$ at the point of collision taking small values of $%
L_{i}$ to keep the right hand side of (\ref{xa}) finite. Then, eqs. (\ref{xd}%
), (\ref{d}) retain their validity. Now, we can neglect $L_{i}$ in (\ref{X}%
), so $X_{i}\approx E_{i}$.

\subsection{Collision in the turning point}

Let near-horizon collision occur in the turning point (say, for particle 1).
Then, $Z_{1}=0.$ Taking for definiteness $L_{1}>0$ we have $\alpha _{1}=%
\frac{\pi }{2}$. $d=1-\sin \alpha _{2}$.

Let this be the turning point for particle 2 as well. Then, if $L_{2}>0$, $%
\alpha _{2}=\frac{\pi }{2}$, $d=0$ in eq. (\ref{d}), the effect of unbounded 
$\gamma $ is absent. If $L_{2}<0$, $\alpha _{2}=-\frac{\pi }{2}$ and $d=2,$%
\begin{equation}
m_{1}m_{2}\gamma \approx \frac{2E_{1}E_{2}}{N^{2}}\text{.}  \label{2t}
\end{equation}

\section{Kinematic explanation}

It is of interest to elucidate the underlying reason that gives rise to
unbounded $\gamma $ for near-horizon collisions in the case (\ref{gf}). For
the standard BSW effect, with $g_{\phi }$ separated from zero, it turned out
that it is due to collision between the slow fine-tuned particle and a rapid
usual one \cite{k}. But now both particles are usual, no fine-tuning is
assumed. Therefore, the explanation of unbounded $\gamma $ should be
different from that given in \cite{k}.

\subsection{Basic kinematic formulas}

To make presentation self-contained, below we repeat and somewhat enlarge
derivation of basic formulas made in \cite{k}. Let us introduce the tetrad
basis $h_{(a)\mu }$ that corresponds to the frame of the zero-angular
momentum observer (ZAMO) \cite{72}. Then,

\begin{equation}
h_{(0)\mu }=-N(1,0,0,0)\text{,}  \label{h0}
\end{equation}%
\begin{equation}
h_{(1)\mu }=\frac{1}{\sqrt{A}}(0,1,0,0)\text{,}  \label{1}
\end{equation}%
\begin{equation}
h_{(2)\mu }=\sqrt{g_{\theta }}(0,0,1,0)\text{,}
\end{equation}%
\begin{equation}
h_{(3)\mu }=\sqrt{g_{\phi }}(-\Omega ,0,0,1)\text{.}  \label{3}
\end{equation}%
Here, $x^{\mu }=(t,r,\theta ,\phi )$. If one identifies $h_{(0)\mu }$ with
the four-velocity $U_{\mu }$ of such an observer, it is seen from (\ref{h0})
that $U_{\phi }=0$, so this does correspond to the ZAMO. The proper time of
the observer under discussion can be obtained according to $d\tau
_{obs}=-dx^{\mu }U_{\mu }$. Then, one can define the tetrad spatial
components of the velocity of a particle:%
\begin{equation}
V^{(i)}=V_{(i)}=\frac{dx^{\mu }h_{(i)\mu }}{d\tau _{obs}}=-\frac{u^{\mu
}h_{(i)\mu }}{u^{\mu }h_{(0)\mu }}\text{.}  \label{V}
\end{equation}%
We apply these formulas to the equatorial motion $\theta =\frac{\pi }{2}$
with $A=N^{2}$. Using (\ref{0}), (\ref{X}) and (\ref{h0}) - (\ref{3}) we
obtain%
\begin{equation}
-u^{\mu }h_{(0)\mu }=\frac{X}{mN}\text{.}
\end{equation}

It follows from (\ref{phi}) and (\ref{3}) that%
\begin{equation}
u^{\mu }h_{(3)\mu }=\frac{L}{m\sqrt{g_{\phi }}}\text{.}
\end{equation}

Eq. (\ref{V}) gives us%
\begin{equation}
V^{(3)}=\frac{LN}{\sqrt{g_{\phi }}X}=\frac{bL}{X}\text{.}  \label{v3}
\end{equation}%
Taking into account (\ref{r}) and (\ref{1}) with $A=N^{2}$, we have the
component in the radial direction%
\begin{equation}
V^{(1)}=\sigma \sqrt{1-\frac{N^{2}}{X^{2}}(m^{2}+\frac{L^{2}}{g_{\phi }})}%
\text{.}  \label{v1}
\end{equation}

Then, calculating the absolute value $V^{2}=\left( V^{(1)}\right)
^{2}+\left( V^{(3)}\right) ^{2}$, we obtain

\begin{equation}
X=\frac{mN}{\sqrt{1-V^{2}}}\text{.}  \label{xn}
\end{equation}%
The gamma factor (\ref{gau}) can be written in the form%
\begin{equation}
\gamma =\frac{1}{\sqrt{1-w^{2}}}\text{,}  \label{gaw}
\end{equation}%
where $w$ has the meaning of relative velocity. Then, 
\begin{equation}
w^{2}=1-\frac{(1-V_{1}^{2})(1-V_{2}^{2})}{(1-\vec{V}_{1}\vec{V}_{2})^{2}}
\label{w}
\end{equation}%
(see, for example, problem 1.3. in \cite{text}).

\subsection{Behavior of a velocity near the horizon}

Now, we will use the above general formulas for the analysis of particles'
motion near the horizon, when $N\rightarrow 0$. For the angle $\beta $
between different components of the velocity we have for small $N$:%
\begin{equation}
\tan \beta =-\frac{V^{(3)}}{V^{(1)}}\approx \frac{bL}{X\sqrt{1-\frac{%
b^{2}L^{2}}{X^{2}}}}\text{.}  \label{beta}
\end{equation}%
Near the horizon, eqs. (\ref{v1}), (\ref{xn}) give us%
\begin{equation}
-V^{(1)}=1-\frac{L^{2}b^{2}}{2X^{2}}+O(N^{2})
\end{equation}%
\begin{equation}
V\approx 1-\frac{N^{2}m^{2}}{2X^{2}},  \label{vn}
\end{equation}%
\begin{equation}
\vec{V}_{1}\vec{V}_{2}\approx (1-\frac{L_{1}^{2}b^{2}}{2X_{1}^{2}})(1-\frac{%
L_{2}^{2}b^{2}}{2X_{2}^{2}})+b^{2}\frac{L_{1}L_{2}}{X_{1}X_{2}}.  \label{v12}
\end{equation}

Thus $V_{1}$ $\rightarrow 1$, $V_{2}\rightarrow 1$. Meanwhile, the mutual
orientation of particles' velocities depends on $b$. Now, we will consider
different cases separately.

\subsection{lim$_{N\rightarrow 0}b\neq 0$}

We see from (\ref{v12}) that

\begin{equation}
\lim_{N\rightarrow 0}\vec{V}_{1}\vec{V}_{2}<1
\end{equation}%
and it follows from (\ref{beta}) that $\beta \neq 0$. Eq. (\ref{w}) entails
that $w\rightarrow 1$, so $\gamma \rightarrow \infty $. Thus the underlying
reason of the effect is that particles approach the horizon under different
angles $\beta $. This is in sharp contrast with the case of nonzero $g_{\phi
}$ when any usual particle hits the horizon perpendicularly with the speed
approaching that of light that leads to $w<1$ \cite{k} and finite $\gamma $
in (\ref{gaw}).

In the case under discussion, both particles approach the horizon with the
speed of light as well, so both of them are rapid. However, because of
vanishing $g_{\phi }$ and nonzero $b$, the projection of, say, velocity of
particle 2 onto the direction of motion of particle 1 is less than 1, so
this piece of motion is slow. Then, we have collision between a rapid
particle and (in this sense) slow one.

There is no need to distinguish case of finite $b$ and $b\rightarrow \infty $
(with $\left\vert bL_{i}\right\vert <X_{i}$ according to (\ref{xa})).
Explanation in question applies to both of them.

\subsection{lim$_{N\rightarrow 0}b=0$}

It follows from (\ref{v12}) that%
\begin{equation}
\vec{V}_{1}\vec{V}_{2}\approx 1-\frac{b^{2}}{2}(\frac{L_{1}}{X_{1}}-\frac{%
L_{2}}{X_{2}})^{2}\text{,}  \label{vb}
\end{equation}%
so the situation is somewhat different. It is seen from (\ref{beta}) that $%
\beta \rightarrow 0$, any particle hits the horizon perpendicularly as in
the standard case \cite{k}. However, there is also a big difference here. It
is seen from (\ref{w}) that whether or not $w\rightarrow 1$ is determined by
the competition of two factors - the rate with which the absolute value of
each particle approaches the speed of light and the rate of mutual alignment
of particles' velocities. In the standard case, both rates are the same.
Indeed, $1-V^{2}=O(N^{2})$ and $\vec{V}_{1}\vec{V}_{2}=1-O(N^{2})$. As a
result, according to (\ref{w}), $w$ is separated from $1$ and the effect of
unbounded $E_{c.m.}$ for two usual particles is absent. Meanwhile, in our
case, because of the small factor $g_{\phi }$ in the expression for $b$,
particles tend to be aligned more slowly. As a result, $w\rightarrow 1$. One
can say that although $\beta \rightarrow 0$, some "memory" about small
nonzero $\beta $ (i.e. direction of motion) still remains.

The role of small $g_{\phi }$ is displayed if one writes the explicit
expression for $w$ near the horizon. It follows from (\ref{w}), (\ref{vn})
and (\ref{vb}) that%
\begin{equation}
w^{2}\approx 1-\frac{4m_{1}^{2}m_{2}^{2}X_{1}^{2}X_{2}^{2}}{%
(L_{1}X_{2}-L_{2}X_{1})^{4}}g_{\phi }^{2}\text{.}  \label{wg}
\end{equation}

It is seen that just the small factor $g_{\phi }$ is responsible for the
result $w\rightarrow 1$. If, instead, we put $g_{\phi }$ separated from
zero, $w$ turns out to be separated from $1$, so the effect of unbounded $%
\gamma $ and $E_{c.m.}$ is absent.

One reservation is in order. There is an exceptional case when $h\equiv
L_{1}X_{2}-L_{2}X_{1}=0$. It follows from (\ref{X}) that $%
h=L_{1}E_{2}-L_{2}E_{1}$ is a constant, so if $h=0$ on the horizon, it
vanishes everywhere. In this case, the main term in the near-horizon
expansion of $\gamma $ having the order $\frac{b^{2}}{N^{2}}=\frac{1}{%
g_{\phi }}$ vanishes. The next terms contain$\frac{b^{4}}{N^{2}}=\frac{N^{2}%
}{g_{\phi }^{2}}$, so for unbounded $\gamma $ it is necessary that $g_{\phi
} $ tend to zero faster than $N$.

\subsection{Finite nonzero versus vanishing $g_{\protect\phi }$: comparison}

It is instructive to summarize the observations made above and compare them
to the standard BSW effect due to particle collision near the horizon with $%
g_{\phi }\neq 0$. In the standard case, one particle should be critical in
the sense that $X_{H}=0$ for it \cite{ban}, \cite{prd}. Taking into account
definition (\ref{X}), one obtains $E=\Omega _{H}L$. Therefore, and one is
led to conclusion that rotation is the necessary ingredient for this
phenomenon. From another hand, for $g_{\phi }\rightarrow 0$, the spacetime
can be static, rotation is not necessary at all since, according to the
explanation above, a small factor $g_{\phi }$ does the job. And, both
particle are usual now, so $X_{H}\neq 0$ for each of them.

This can be displayed in Table 1.

\begin{tabular}{|l|l|l|}
\hline
& $g_{\phi }$ $\neq 0$ on horizon & $g_{\phi }=0$ on horizon \\ \hline
Rotation & necessary & unnecessary \\ \hline
Particles & one critical and one usual & two usual \\ \hline
\end{tabular}

Table 1. Comparison of conditions that lead to unbounded $E_{c.m.}$
depending on $g_{\phi }$ on the horizon.

\section{Examples}

In this section, we show that several particular models considered before
fall in the class under discussion.

\subsection{Stringy black hole}

In the dilaton gravity with the electromagnetic field, the following exact
solution \cite{bron}, \cite{str} is known:

\begin{equation}
ds^{2}=-(1-\frac{2M}{r})dt^{2}+\frac{dr^{2}}{1-\frac{2M}{r}}%
+r(r-r_{+})(d\theta ^{2}+\sin ^{2}\theta d\phi )^{2}\text{.}  \label{s}
\end{equation}%
Here $r_{+}$ is the horizon radius,%
\begin{equation}
r_{+}=\frac{Q^{2}}{M}e^{-2\varphi _{0}}\text{,}
\end{equation}%
$Q$ is the electric charge, $M$ being mass, $\varphi _{0}$ is the dilaton
field at infinity.

Let a black hole be extremal,%
\begin{equation}
Q^{2}=2M^{2}e^{2\varphi _{0}}.
\end{equation}

Then,%
\begin{equation}
r_{+}=2M\text{,}
\end{equation}%
\begin{equation}
\lim_{r\rightarrow r_{+}}b=\frac{1}{r_{+}}.
\end{equation}

In doing so, the horizon area $A=4\pi \lim_{r\rightarrow r_{+}}r(r-r_{+})=0$%
, the surface shrinks to the point.

In \cite{fern}, collision with participation of charged particles was
considered. However, this is not necessary since the effect of unbounded $%
\gamma $ persists even for collision of neutral ones. Now, eq. (\ref{gd})
applies and we obtain unbounded $\gamma $ in agreement with \cite{fern}.

\subsection{Black hole in Brans-Dicke theory}

In the Brans-Dicke theory an exact solution is known (see \cite{kim} and
references therein):

\begin{equation}
ds^{2}=\Delta ^{-\frac{2}{2\omega +3}}\sin ^{-\frac{4}{2\omega +3}}\theta
\lbrack -dt^{2}(1-\frac{2M}{r})+r^{2}\sin ^{2}\theta d\phi ^{2}]+\Delta ^{%
\frac{2}{2\omega +3}}\sin ^{\frac{4}{2\omega +3}}\theta (\frac{dr^{2}}{1-%
\frac{2M}{r}}+r^{2}d\theta ^{2})\text{.}
\end{equation}%
Here, $\omega $ is the parameter of the Brans-Dicke theory, $\Delta =r(r-2M)$%
, the horizon lies at $r_{+}=2M$. This metric represents the Brans-Dicke
counterpart of the Schwarzschild metric (denoted for shortness as the BDS
metric).

Now, 
\begin{equation}
\left( N^{2}\right) =\frac{\Delta ^{\frac{2\omega +1}{2\omega +3}}}{r^{2}}%
\text{,}
\end{equation}%
\begin{equation}
g_{\phi }=\Delta ^{-\frac{2}{2\omega +3}}r^{2}\text{,}
\end{equation}%
where we put $\theta =\frac{\pi }{2}$,%
\begin{equation}
b^{2}=\frac{\Delta }{r^{4}}=\frac{r-r_{+}}{r^{3}}\text{.}
\end{equation}

Curvature invariants are finite at $r=r_{+}$ for $-\frac{5}{2}\leq \omega <-%
\frac{3}{2}$ \cite{kim}. In this interval, $2\omega +3<0$, and $g_{\phi
}\rightarrow 0.$ Simultaneously, $g_{\theta }\rightarrow \infty $ that
compensates the small factor $g_{\phi }$, the horizon area remains finite, $%
A=4\pi ^{2}r_{+}$. Now, $\lim_{r\rightarrow r_{+}}b=0$, so Eq. (\ref{gc})
applies.

In \cite{s2}, the rotating counterpart of the Kerr metric in the Brans-Dicke
theory (so-called the BDK metric) was considered. The result $\gamma
\rightarrow \infty $ was obtained. Meanwhile, we see that this effect
persists even in the BDS metric, so rotation is not necessary here.

\section{Janis-Newman-Winicour metric}

This metric \cite{jnwmet} - \cite{form} can be written in the form

\begin{equation}
ds^{2}=-(1-\frac{r_{+}}{r})^{\nu }dt^{2}+\frac{dr^{2}}{(1-\frac{r_{+}}{r}%
)^{\nu }}+r^{2}(1-\frac{r_{+}}{r})^{1-\nu }(d\theta ^{2}+d\phi ^{2}\sin
^{2}\theta ).
\end{equation}%
Now, $N^{2}=(1-\frac{r_{+}}{r})^{\nu }$, $g_{\phi }=r^{2}(1-\frac{r_{+}}{r}%
)^{1-\nu },$ $\lim_{r\rightarrow r_{+}}g_{\phi }=0$, provided $\nu <1$. Thus
for $r=r_{+}$ the sphere of $\ $a constant $r$ shrinks to the point and
represents a singular point rather than a regular black hole horizon. In
doing so,%
\begin{equation}
b^{2}=\frac{1}{r^{2}}(1-\frac{r_{+}}{r})^{2\nu -1}.
\end{equation}

To handle the limit $N\rightarrow 0$, our approach applies. Three different
cases can be considered separately. Following \cite{jnw}, we assume that $%
E_{1}=E_{2}=m$.

\subsubsection{$\protect\nu <\frac{1}{2}$}

In this case, $b\rightarrow \infty $ when $r\rightarrow r_{+}$. If,
additionally, we assume that collision occurs in the common turning point
for particles 1 and 2 having angular momenta of different signs we obtain
from (\ref{cm}), (\ref{2t}) that%
\begin{equation}
E_{c.m.}^{2}\approx \frac{4m^{2}}{(1-\frac{r_{+}}{r})^{\nu }}\text{.}
\end{equation}%
This coincides with eq. (29) of \cite{jnw}.

$\nu =\frac{1}{2}$, $b(r_{+})=\frac{1}{r_{+}^{2}}\neq 0$

Now, we use (\ref{cm}), (\ref{gd}). This corresponds just to eq. (31) of 
\cite{jnw}.

\subsubsection{$\protect\nu >\frac{1}{2}$}

Now, $b(r_{+})=0$, (\ref{c}) gives us%
\begin{equation}
C=\frac{(L_{1}-L_{2})^{2}}{2}\text{.}
\end{equation}%
Then, we have from (\ref{cm}) and (\ref{gc}) that%
\begin{equation}
E_{c.m.}^{2}\approx \frac{(L_{1}-L_{2})^{2}}{r_{+}^{2}(1-\frac{r_{+}}{r}%
)^{1-\nu }}
\end{equation}%
that agrees with eq. (32) of \cite{jnw} (the extra factor 1/2 in \cite{jnw}
is an obvious typo in passing from their eq. 31 to eq. 32).

\subsection{Geometry versus material source}

In \cite{jnw}, the authors attributed indefinitely high $E_{c.m.}$ to the
presence of the scalar field. This was criticized in \cite{s2} where the
main emphasis was made on the role of interaction between particles and the
scalar field that can lead to diminishing $E_{c.m.}$. However, the result of
infinite $E_{c.m.}$ for collision of free particles in the BDK metric
remained in \cite{s2} without explanation. Now we see that it is the key
property (\ref{gf}) which is responsible for the effect under discussion.
Moreover, the material source that supports such a metric can include other
fields than the scalar one. Say, the essential ingredient in the case of
metric (\ref{s}) is the presence of the electromagnetic field along with the
scalar one.

\section{Summary}

Thus we developed a general approach to acceleration of particles due to
near-horizon collisions in the background of axially symmetric rotating
black holes, when $g_{\phi }=0$ on the horizon. In the invariant form this
can be written as $\eta _{\mu }\eta ^{\mu }=0$, where $\eta ^{\mu }$ is the
Killing vector responsible for rotation. It turned out that, in contrast to
the standard case with $g_{\phi }\neq 0$, now the effect of unbounded $%
E_{c.m.}$ is obtained for collision of any two usual (not fine-tuned)
particles. The kinematic explanation is found. In contrast to the standard
case \cite{k} where a rapid particle hits the slow one near the horizon, now
both particles are rapid. The effect is achieved due to the angle between
particles.

It is clear from derivation that the source of high $E_{c.m.}$ is not
related to the scalar field or any other concrete source. The main factor is
the geometry. If condition (\ref{gf}) is satisfied, the effect persists both
for naked singularities and regular black hole horizons. If the horizon is
regular, vanishing $g_{\phi }$ is compensated by high $g_{\theta }$. In this
sense, the effect near regular horizons is connected with their high degree
of anisotropy. For naked singularities this is not necessary, the isotropic
case is also suitable.

The effect exists both for rotating and static $(\Omega =0$) metrics. In
this sense, rotation is not necessary in contrast to the standard BSW
effect. Also, there is no need for the electric charge of particles even for
static metrics in contrast to the standard case \cite{jl}. The effect under
discussion reveals itself even for collision of usual particles, so
fine-tuning is not required.

Thus the standard BSW effect and the one considered in the present work give
complete unified picture, being its different realizations.

It would be of interest to extend the results of this work further relaxing
the condition of axial symmetry.

\begin{acknowledgments}
This work was funded by the subsidy allocated to Kazan Federal University
for the state assignment in the sphere of scientific activities.
\end{acknowledgments}


\begin{thebibliography}{99}
\bibitem{ban} M. Ba\~{n}ados, J. Silk and S.M. West, Kerr black Holes as
particle accelerators to arbitrarily high Energy, Phys. Rev. Lett. \textbf{%
103} (2009) 111102 [arXiv:0909.0169].

\bibitem{prd} O.B. Zaslavskii, Acceleration of particles as universal
property of rotating black holes, Phys. Rev. D \textbf{82} (2010) 083004
[arXiv:1007.3678].

\bibitem{jnw} M. Patil and P. Joshi, Acceleration of particles by
Janis-Newman-Winicour singularities, Phys.\ Rev. D \textbf{85}, 104014
(2012).

\bibitem{fern} S. Fernando, String black hole: can it be a particle
accelerator ?, Gen. Relativ. Gravit. \textbf{46,} 1634 (2014), [
arXiv:1311.1455].

\bibitem{s2} J. Sultana and B. Bose, Particle collisions near a Kerr-like
black hole in Brans-Dicke theory, Phys. Rev. D \textbf{92}, 104022 (2015).

\bibitem{kim} H. Kim, New black hole solutions in Brans-Dicke theory of
gravity, Phys. Rev. D \textbf{60}, 024001 (1999) [arXiv:gr-qc/9811012].

\bibitem{dirty} I. V. Tanatarov and O. B. Zaslavskii, Dirty rotating black
holes: regularity conditions on stationary horizons, Phys. Rev. D \textbf{86}%
, 044019 (2012) [arXiv:1206.2580].

\bibitem{k} O. B. Zaslavskii, Acceleration of particles by black holes:
kinematic explanation, Phys. Rev\textit{.} D \textbf{84}, 024007 (2011)
[arXiv:1104.4802].

\bibitem{72} J. M. Bardeen, W. H. Press, and S. A. Teukolsky, Rotating black
holes: locally nonrotating frames, energy extraction and scalar synchrotron
radiation, Astrophys. J. \textbf{178}, 347 (1972).

\bibitem{text} A. P. Lightman, W. H. Press, R. H. Price, and S. A.
Teukolsky, \textit{Problem book in Relativity and Gravitation} (Princeton
University Press, Princeton, New Jersey, 1975).

\bibitem{bron} K. A. Bronnikov and G. N. Shikin, Interacting fields in
general relativity theory, Izv. Vuzov SSSR Fiz. \textbf{9}, 25 (1977) (Russ.
Phys. J. \textbf{20}, 1138 (1977)).

\bibitem{str} Garfinkle, D., Horowitz, G.T., Strominger, A.: Charged black
holes in string theory. Phys. Rev. D 43, 3140 (1991) Erratum 45 3888 (1992).

\bibitem{jnwmet} A. I. Janis, E. T. Newman, and J. Winicour, Reality of the
Schwarzschild singularity, Phys. Rev. Lett. \textbf{20}, 878 (1968).

\bibitem{wyn} M. Wyman, Static spherically symmetric scalar fields in
general relativity, Phys. Rev. D \textbf{24}, 839 (1981).

\bibitem{form} K. S. Virbhadra, S. Jhingan, and P. S. Joshi, Nature of
singularity in Einstein-Massless scalar theory, Int. J. Mod. Phys. D \textbf{%
6}, 357 (1997).

\bibitem{jl} O. Zaslavskii, Acceleration of particles by nonrotating charged
black holes, Pis'ma ZhETF \textbf{92}, 635 (2010) (JETP Letters \textbf{9}2,
571 (2010)), [arXiv:1007.4598].
\end{thebibliography}
\end{document}